\documentclass[prl,
a4paper,
nofootinbib,
twocolumn,
reprint,
]{revtex4}
\usepackage{graphicx}
\usepackage{amsfonts}
\usepackage{amssymb}
\usepackage{slashed} 
\usepackage{color}
\usepackage{hyperref}
\def\eq#1{{Eq.~(\ref{#1})}}
\def\eqs#1#2{{Eqs.~(\ref{#1})--(\ref{#2})}}

\def\Im{\mbox{Im}\,}
\def\Re{\mbox{Re}\,}

\definecolor{oucrimsonred}{rgb}{0.6, 0.0, 0.0}
\definecolor{persianblue}{rgb}{0.11, 0.22, 0.73}
\definecolor{forestgreen}{rgb}{0.13,0.35,0.13}
 \hypersetup{colorlinks, citecolor=oucrimsonred, linkcolor=persianblue, urlcolor=oucrimsonred}
\def\hhref#1{\href{http://arxiv.org/abs/#1}{#1}} 
\newcommand{\be}{\begin{equation}}
\newcommand{\ee}{\end{equation}}
\newcommand{\bea}{\begin{eqnarray}}
\newcommand{\eea}{\end{eqnarray}}
\newcommand{\nn}{\nonumber}

\begin{document}
\title[]{$Z$ Boson Decay into Light and Darkness
}
\date{\today}
\author{M.\ Fabbrichesi$^{\dag}$}
\author{E.\ Gabrielli$^{\ddag\dag}$}
\author{B.\ Mele$^{\ast}$}
\affiliation{$^{\dag}$INFN, Sezione di Trieste, Via  Valerio 2, 34127 Trieste, Italy }
\affiliation{$^{\ddag}$Physics Department, University of Trieste, Trieste, Italy}
\affiliation{$^{\ddag}$ NICPB, R\"avala 10, Tallinn 10143, Estonia }
\affiliation{$^{\ast}$INFN, Sezione di Roma, P.le Aldo Moro 2, 00185 Roma, Italy}
\begin{abstract}
\noindent  
We study the  $Z\rightarrow \gamma \bar \gamma$   process in which the $Z$ boson decays into a  photon $\gamma$  and a  massless dark photon $\bar \gamma$,  when the  latter couples to standard-model fermions via dipole moments. 
This is a simple yet nontrivial example of how the Landau-Yang theorem---ruling out the decay of a massive spin-1 particle into two  photons---is evaded if the final particles can be distinguished.
The striking signature of this process is a resonant monochromatic single photon in the $Z$-boson center of mass  together with missing momentum.
LEP  experimental bounds allow a branching ratio up to about 10$^{-6}$ for such a decay. In a simplified model of  the dark sector, the dark-photon dipole moments
 arise from one-loop exchange of heavy dark fermions and scalar messengers.
The corresponding  prediction for the rare $Z\to \gamma \bar \gamma$ decay width can be  explored with the large samples
of $Z$ bosons foreseen at future colliders.

\end{abstract}
\maketitle

Consider the decay of a massive spin-one particle into two  massless spin-one particles. At first glance, this channel should vanish---as it does in the case of two final photons---as dictated by the Landau-Yang theorem~\cite{Landau:1948kw}. Yet the theorem need not apply
if the two final states can be distinguished. This is the case when the final state is made of a  photon $\gamma$ and  a  \textit{dark} photon $\bar \gamma$. 

The possibility of extra  $U(1)$ gauge groups---with  dark photons mediating interactions among the dark-sector particles, which are uncharged under the 
standard-model (SM) gauge groups---is the subject of many theoretical speculations and experimental searches (see Ref.~\cite{Raggi:2015yfk} for recent reviews, mostly for the massive case).  

The case of \textit{massless} dark photons is  perhaps the most interesting  because the dark photon can be completely decoupled from the SM~\cite{Holdom:1985ag}, and 
interactions between SM fermions and dark photons take place only by means of higher-order operators~\cite{Dobrescu:2004wz}, which are  automatically  suppressed.  Possible experimental tests of this scenario have been investigated in Higgs physics~\cite{Biswas:2016jsh}, flavor-changing neutral currents~\cite{Gabrielli:2016cut}, and kaon physics~\cite{Fabbrichesi:2017vma}. Its relevance for dark-matter  dynamics has been discussed in Ref.~\cite{Foot:2004pa}.

The decay of a $Z$ boson into one SM and one dark photon would be a most striking signature for both the existence of  dark photons, and the embodiment of the nonapplicability of  the  Landau-Yang theorem. The process can proceed at one loop via 
SM-fermion exchange. To bypass the theorem, the photon and dark photon must couple differently  to the fermions in the loop so as to be distinguishable. This naturally occurs  for {\it massless} dark photons since they do not have a Dirac  (i.e. mediated by a single $\gamma$ matrix) interaction  but only a Pauli (i.e. mediated by two $\gamma$ matrices) dipole interaction:
\bea
{\cal L}\sim \bar{\psi}\, \sigma_{\mu\nu} \left( d_M + i \gamma_5\, d_E \right)\psi  \label{first}
\,B^{\mu\nu}\, ,
\eea
where  $B_{\mu\nu}$ is  the  field strength associated with the dark photon field $B_{\mu}$,
and $\sigma_{\mu\nu}=1/2[\gamma_{\mu},\gamma_{\nu}]$.

For {\it massive} dark photons, $z'$, the leading interaction  would be of the same 
SM-photon Dirac type as the photon,  
$ e\, \bar{\psi}\, \gamma_{\mu} \psi B^{\mu}$,
decreased by the  mixing parameter $\epsilon$~\cite{Raggi:2015yfk}.  The $Z\rightarrow \gamma\, z'$ channel would then be  doubly suppressed by an $\epsilon^2$  factor and  an extra
term ${\cal O} (m_{z'}^2/M_Z^2)$, which brings back the outcome of the Landau-Yang theorem for $m_{z'}\to 0$.
In this case too then, the higher-order Pauli dipole interaction might be the most relevant, as it is in the case of a massless dark photon. The following analysis can then be extended in a straightforward way to the 
massive dark-photon case.

The  experimental signature for  $Z\rightarrow \gamma \bar \gamma$  is  quite simple and distinctive. In the $Z$-boson center-of-mass  frame, the photon is monochromatic with an energy of about 45 GeV. A massless dark photon 
has a neutrinolike signature in a typical experiment~\cite{Biswas:2016jsh}, and appears as missing momentum in the $Z\to \gamma + X$ final state.  Such a process has been  explored at  LEP (in the assumption of $X$ being either a $\nu \bar \nu$ pair or   a hypothetical axion, if sufficiently light) to find the limit  of $10^{-6}$ (95\% C.L.) for the corresponding branching ratio (BR)~\cite{Acciarri:1997im}.  

\vskip0.3em
\textit{Effective dipole moments in a simplified model of the dark sector.}---We compute the dipole operators of \eq{first} in
 a simplified-model framework, where we make as few assumptions as possible on the structure of the dark sector. 

We extend the SM field content by 
a  new (heavy) dark  fermion $Q$, which is a singlet under the SM gauge interactions, but is charged under the unbroken $U_D(1)$ gauge group associated to the massless dark photon. We focus on the up-quark kind of interaction. 
The dark fermion couples to  SM fermions by means of a Yukawa-like interaction given by 
\be
{\cal L} \, \, \supset \, \,  g^f_L (\bar{Q}_L q^f_R) S_R + g^f_R (\bar{Q}_R q^f_L) S_L + H.c. \; ,\label{lag}
\ee
where $S_L$ and $S_R$ are new (heavy) {\it messenger} scalar particles, and $S_L$ is an  
$SU(2)$ doublet.  In \eq{lag},   $q^f_L$ and $q^f_R$ stand for  SM fermions of flavor $f$---that is, $SU(3)$ triplets   and, respectively, $SU(2)$ doublets and singlets.  The $S_L$ messenger  field   is a $SU(2)$ doublet,  $S_R$   is a $SU(2)$ singlet, and both are $SU(3)$ color triplets (singlets) for quark (lepton) messengers. Both fields are also charged under $U_D(1)$, carrying the same dark-fermion charge.

In order to generate chirality-changing processes, the mixing terms of the kind
(see Ref.~\cite{Gabrielli:2013jka} for more details)
 \be
{\cal L} \, \, \supset \, \, \lambda_S S_0 \left( \tilde H^\dag  S_L  S_R^{\dag}  +  S_L^{\dag} S_R H \right) \, , \label{mix} 
\ee
are required, 
where $H$ is the SM Higgs boson, $\tilde{H}=i\sigma_2 H^\ast$, and $S_0$ a scalar singlet. 
After both   $S_0$ and $H$   take a vacuum expectation value (VEV)  ($\mu_S$  and $v$---the electroweak VEV---respectively), the Lagrangian in \eq{mix} gives rise to  the mixing. 

Then, each of the messenger fields $S_\pm$ (obtained from $S_{L,R}$
by diagonalization)  
couples  to both left- and right-handed SM fermions of flavor $f$ with strength $g^f_L/\sqrt{2}$  and $g^f_R/\sqrt{2}$, respectively. We can assume that the size of the  mixing [proportional to the product  of the VEVs  ($\mu_s v$)] is large and of the same order  of the heavy-fermion and heavy-scalar masses.

The resulting model
can be considered as a template for many models of the dark sector, and is 
a simplified version of the model in Ref.~\cite{Gabrielli:2013jka}, which might  provide a natural solution to the SM flavor-hierarchy problem.

The SM Lagrangian plus the  terms in \eqs{lag}{mix}  and the corresponding kinetic terms provides a simplified model for the dark sector and the effective interaction of a  massless dark photon $\bar\gamma$ with the SM fields. 
Then, SM fermions couple to $\bar\gamma$ only via non-renormalizable interactions~\cite{Dobrescu:2004wz} induced by loops of  dark-sector  particles. 
The corresponding effective Lagrangian is equal to 
\be
{\cal L}=\sum_f \frac{e_D}{2\Lambda} \bar{\psi}_f \sigma_{\mu\nu} \left( d^f_M + i \gamma_5 d^f_E \right)\psi_f \label{dipole}
B^{\mu\nu}\, ,
\ee
where the sum runs over all the SM fields, $e_D$ is the $U_D(1)$ dark elementary charge (we assume universal couplings), $\Lambda$ the  effective scale of the dark sector, $\psi_f$ a generic SM fermion field. The magnetic  and electric dipoles are given by 
\be
d^f_M =\frac{1}{2}  \Re \frac{g^f_L g^{f*}_R}{(4 \pi)^2} \quad \mbox{and}  \quad d^f_E = \frac{1}{2}  \Im \frac{g^f_L g^{f*}_R}{(4 \pi)^2}  \, , \label{ds}
\ee
respectively.

 \begin{figure}[t]
\begin{center}
\includegraphics[width=3.4in]{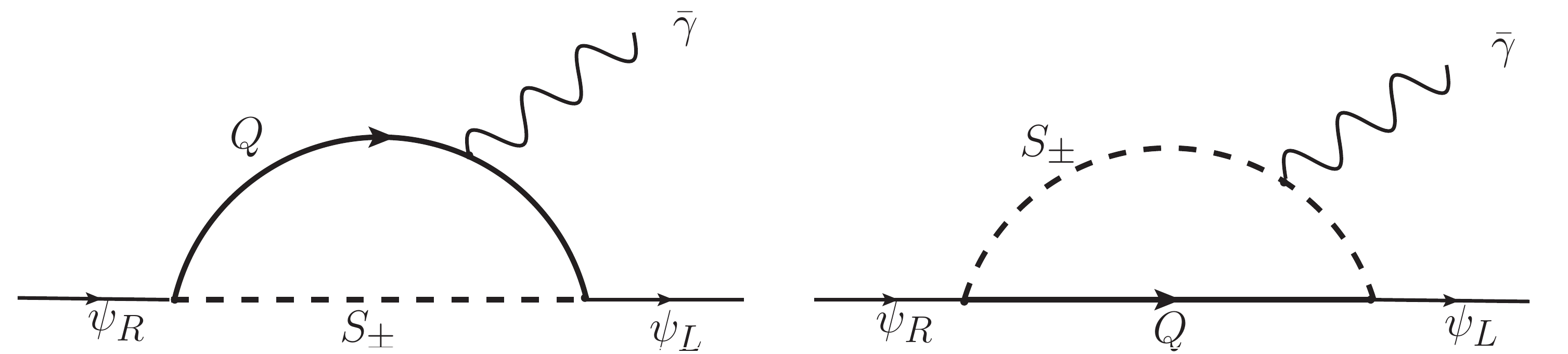}
\caption{\small One-loop vertex diagrams giving rise to the effective  dipole operators  in~\eq{dipole} between SM fermions and the dark photon $\bar \gamma$.  Dark-sector fermions ($Q$) and  scalars ($S_\pm$) run inside the loop. $\psi_L$ and $\psi_R$ are  SM chiral fermions of  arbitrary flavor. 
\label{loop} }
\end{center}
\end{figure}

The operators in   \eq{dipole} arise via one-loop diagrams after integrating out the heavy 
dark-sector states (see Fig.~\ref{loop}).
Two mass parameters are relevant in the integration: the dark-fermion mass $M_Q$,  parametrizing  chiral-symmetry breaking in the dark sector, and the mass of the lightest-messenger  $m_S$. As far as the contribution to the magnetic-dipole operator (with vanishing 
quark masses) is concerned,  
for $m_S \gg M_Q$ one has a chiral suppression, with a  $M_Q/m_S^2$ scaling, 
while  
for $m_S \ll M_Q$  one has a $1/M_Q$  behavior, due to the decoupling built in the theory. 
In order to reduce the number of  dimensionful parameters, we have introduced 
in   \eq{dipole}
a dark-sector effective scale $\Lambda$, defined as  
 the common mass of the  dark fermion and the lightest messenger scalar. This choice    corresponds  to the maximal chiral enhancement.
 Nevertheless, because  the dipole moment in \eq{dipole} is  proportional to the  
 messenger mixing [see~\eq{mix} and following text], the effective scale 1/$\Lambda$ is also proportional to the ratio $v \mu_S/m_S^2$, as expected from the $SU(2)$ symmetry breaking. Since we are assuming a large-mixing scenario,  $v \mu_S/m_S^2$ is of order 1, and one can express the effective scale as in  \eq{dipole} with  $\Lambda \sim m_S~\sim M_Q $.

Stringent  limits on the scale and couplings of the dark sector come from flavor physics~\cite{Gabrielli:2016cut,Fabbrichesi:2017vma} and astrophysics~\cite{Hoffmann:1987et}. In order to evade them, we restrict ourselves to flavor diagonal interactions of heavier  quarks and leptons for which there are currently no bounds.

\begin{figure}[t]
\begin{center}
\includegraphics[width=3.4in]{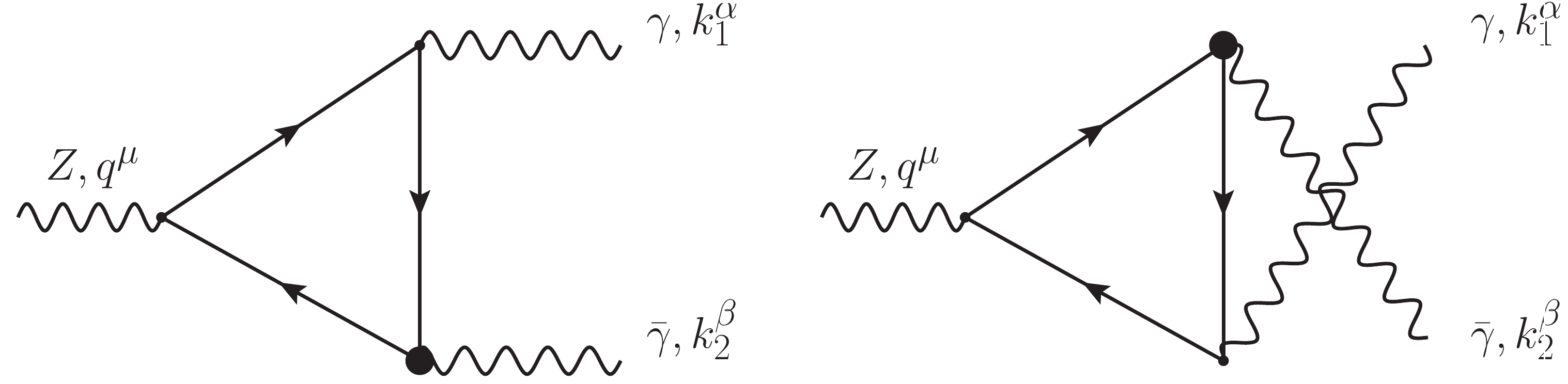}
\caption{\small Feynman diagrams for the decay $Z(q) \to \gamma(k_1) \bar{\gamma}(k_2)$. The blob represents the insertion of the dipole operator in \eq{dipole}. The case of two photons with the same interaction (no blobs) would lead to a cancellation as dictated by the Landau-Yang theorem.
\label{Zdecay} }
\end{center}
\end{figure}

\vskip0.3em
\textit{Amplitudes.}---We are interested in the decay process of a $Z$ boson into two massless spin-one particles:
\be 
Z(q) \to  \gamma(k_1) \bar{\gamma}(k_2)\, ,
\ee 
where $k_1$ and $k_2$ are the photon and dark-photon 4-momenta, respectively, and $q=k_1+k_2$ is  the $Z$-boson 4-momentum. The total amplitude $\cal M$ for the decay process is obtained by computing the one-loop diagrams represented in Fig.~\ref{Zdecay}. It is given by
\be
{\cal M} ={\cal M}_M + {\cal M}_E \label{amp0}
\ee
where the ${\cal M}_M$  and $ {\cal M}_E$ stand for the magnetic- and electric-dipole moment contributions. 

In both  amplitudes in \eq{amp0} the ultraviolet divergencies cancel out in the sum of the two diagrams in Fig.~\ref{Zdecay}, and the result  is  finite. 
We use dimensional regularization. The $\gamma_5$ matrix can be treated naively as anticommuting with all other $\gamma$ matrices  as long as the final expression  is  fixed by requiring that  the Ward identities are satisfied~\cite{Adler:1969gk}. 
All terms are proportional to the fermion masses and no gauge anomaly is involved.

The $CP$-conserving part of the amplitude  is given by
\be
{\cal M}_M =\frac{e_{\scriptscriptstyle D} e}{4\pi^2}\sum_{i=1}^3 C_i Q_i^{\mu\alpha\beta} \,
\epsilon^{\scriptscriptstyle Z}_{\mu}(q)\epsilon_{\alpha}(k_1)\bar{\epsilon}_{\beta}(k_2) \label{amp}
\ee
where $e$ is the electric charge, $\epsilon^Z_{\mu}(q)$, $\epsilon_{\alpha}(k_1)$, $\epsilon_{\beta}(k_2)$ are the $Z$, $\gamma$, and $\bar{\gamma}$ polarization vectors,   respectively, and  
\bea
Q_1^{\mu\alpha\beta}&=&\varepsilon^{\alpha \beta \mu \lambda} k_{1\lambda}-
\frac{2}{M_Z^2}\varepsilon^{\alpha \mu \lambda \rho }  k_1^{\beta} k_{1\lambda} k_{2\rho}
\\
Q_2^{\mu\alpha\beta}&=&\varepsilon^{\alpha \beta \mu \lambda} k_{2\lambda}-
\frac{2}{M_Z^2}\varepsilon^{\beta \mu \lambda \rho }  k_2^{\alpha} k_{1\lambda} k_{2\rho}
\\
Q_3^{\mu\alpha\beta}&=&\frac{1}{M_Z^2}\left(k_1^{\mu}-k_2^{\mu}\right)\varepsilon^{\alpha \beta \lambda \rho } k_{1\lambda} k_{2\rho}
\eea
are gauge invariant operators. The coefficients $C_i$ are given by
\bea
C_1&=&\sum_f \frac{ r_f m_f d_M^f}{\Lambda}\Big( 2+B_f+2 C_f M^2_Z\Big)\nn  \\
C_2&=& \sum_f \frac{ r_f m_f d_M^f}{\Lambda}\Big( 3+2B_f-2 C_f m^2_f\Big)\\
C_3&=&\sum_f \frac{r_f m_f d_M^f}{\Lambda}\Big(  11+5B_f+2C_f (m^2_f+M^2_Z)\Big) \nn \, .
\eea
where $r_f = N_c^f g_A^f  Q_f $.
The  sum  runs over all charged  SM fermions $f$, with $m_f$ the SM fermion masses. Furthermore, $g^f_A=g T_3^f /(2\cos{\theta_W})$ is  the $Z$-boson axial coupling to SM fermions, with $g$ the weak coupling, $\theta_W$ the Weinberg angle, and $T^3_f(=\pm 1/2)$ the eigenvalue of the third component of weak isospin, $N_c^f=3\, (1)$ for quarks (leptons), and $Q_f$ is the electric charge in units of the elementary charge $e$. The $B_f$ and $C_f$ terms 
are defined as
\bea
B_f&\equiv &{\rm Disc}[B_0(M_Z^2, m_f, m_f)], \nn \\
C_f&\equiv & C_0(0, 0, M_Z^2, m_f, m_f, m_f)\, ,
\eea
with $B_0$ and $C_0$ the scalar two- and three-point Passarino-Veltman functions, respectively (see Ref.~\cite{Passarino:1978jh} for their explicit expressions), and ${\rm Disc}[B_0]$ the discontinuity of the function. They  are both finite functions that can be   evaluated numerically, for example, by {\tt P{\footnotesize ACKAGE}~X}~\cite{Patel:2015tea}. 

Then one has
\be
\frac{1}{3} \sum_{pol} {\cal M}_M {\cal M}^\dag_M = \frac{2}{3} \frac{\alpha_{\scriptscriptstyle D}\alpha }{\pi} M_Z^2 |C_M|^2 \, , \label{sumM}
\ee
where $\alpha_D=e_D^2/4\pi$ and $\alpha=e^2/4\pi$ are the fine structure constants, and $C_M = \sum_f d_M^f \, \xi^f (m_f)$, where
\be
\xi^f(m_f) \equiv \frac{r_f m_f }{\Lambda}\Big(3 +B_f+2m^2_fC_f\Big)\, . \label{CM}
\ee

The $CP$-violating contribution to the on-shell amplitude induced by the electric-dipole moment in \eq{dipole} is  given by
\be
{\cal M}_E =i\frac{e_{\scriptscriptstyle D} e}{4\pi^2} C_E\, (k_1^{\mu}-k_2^{\mu})g^{\alpha\beta}
\epsilon^{\scriptscriptstyle Z}_{\mu}(q)\epsilon_{\alpha}(k_1)\bar{\epsilon}_{\beta}(k_2) \, .
\label{AmpCP}
\ee
Accordingly, we find that
\be
\frac{1}{3} \sum_{pol} {\cal M}_E {\cal M}^\dag_E = \frac{2}{3} \frac{\alpha_{\scriptscriptstyle D}\alpha }{\pi} M_Z^2 |C_E|^2 \, , \label{sumE}
\ee 
where $C_E= \sum_f d^f_E \,\xi^f(m_f) $.

The amplitudes in \eq{amp} and \eq{AmpCP} are both proportional to the $Z$-boson axial coupling $g_A$.
In the limit of restored $SU(2)$ symmetry, both squared amplitudes go to zero as $M_Z \rightarrow 0$.

In the on-shell amplitude, all polarization vectors satisfy the transversality condition, namely $\epsilon_{\mu}(k)k^\mu=0$, with $\epsilon_{\mu}$ a generic polarization vector. One can  verify that the amplitudes in \eq{amp} and \eq{AmpCP} satisfy the Ward identities by substituting the polarizations $\epsilon_{\alpha}(k_1)$ and $\epsilon_{\beta}(k_2)$  with the corresponding momenta. 

For the $CP$-conserving part, the Ward identity for  the $Z$ boson---obtained by substituting  $\epsilon^{\scriptscriptstyle Z}_{\mu}(q)$ with $q_{\mu}$---requires a [$SU(2)$ invariant]  counterterm  $HH^\dag F_{\mu\nu}\bar F^{\mu\nu}$  in the effective theory because of the divergence generated by the insertion of the dipole operator in the diagram where the $Z$ Goldstone boson decays. This term does not affect our computation.

\vskip0.3em 
\textit{Lagrangians.}---It is useful to see how the above amplitudes can be derived from a manifestly gauge invariant Lagrangian in  configuration space. In particular, for the Lagrangian induced by the magnetic-dipole moment, we have
\bea
{\cal L}^{(M)}_{eff}= \frac{e_{\scriptscriptstyle D} e}{4\pi^2 M_Z^2}\sum_{i=1}^3 \bar{C}_i {\cal O}_i(x) \, ,
\label{LeffMD}
\eea
where the dimension-six operators ${\cal O}_i$ are given by
\bea
{\cal O}_1 (x )&=&Z_{\mu\nu}\tilde{B}^{\mu\alpha} A^{\nu}_{~\alpha} \, ,\\
{\cal O}_2 (x) &=&Z_{\mu\nu}B^{\mu\alpha} \tilde{A}^{\nu}_{~\alpha} \, ,\\  
{\cal O}_3  (x) &=& \tilde{Z}_{\mu\nu}B^{\mu\alpha} A^{\nu}_{~\alpha}  \, .
\eea
The field strengths $F_{\mu\nu}\equiv\partial_{\mu}F_{\nu} -\partial_{\nu}F_{\mu}$, for $F_{\mu\nu}=(Z,B,A)_{\mu\nu}$, correspond to the $Z$ boson ($Z_{\mu}$), dark photon ($B_{\mu}$), and photon ($A_{\mu}$)  fields, respectively, and 
$\tilde{F}^{\mu\nu}\equiv \varepsilon^{\mu\nu\alpha\beta}F_{\alpha\beta}$ is the dual field strength. Matching the on-shell amplitude for the 
$Z\to \gamma \bar{\gamma}$ process---as obtained by using the effective Lagrangian in \eq{LeffMD}---with the corresponding one in \eq{amp}, yields 
\bea
\bar{C}_1&=&-\sum_f \frac{ r_f m_f d_M^f}{\Lambda}\Big( 5+2B_f+2C_f\left(m_f^2+M_Z^2\right)\Big)\, ,\nn \\
\bar{C}_2&=&-3\sum_f \frac{ r_f m_f d_M^f}{\Lambda}\Big(2+B_f\Big)\, , \nn \\
\bar{C}_3&=&2\sum_f \frac{ r_f m_f d_M^f}{\Lambda}
\Big(4+2B_f+C_fM_Z^2\Big)\, .
\eea

The  Lagrangian induced by the electric-dipole moment is instead 
\bea
{\cal L}^{(E)}_{eff}= \frac{e_{\scriptscriptstyle D} e}{4\pi^2 M_Z^2}C_E {\cal O}(x)  \, ,
\label{LeffED}
\eea
where $C_E$ is given below \eq{sumE} and the dimension-six operator is 
\bea
{\cal O} (x)=Z_{\mu\nu} A^{\mu\alpha} B^{\nu}_{~\alpha} \, .
\eea

The local operators entering the  Lagrangians in 
Eqs.(\ref{LeffMD}) and (\ref{LeffED}) are both $C$ odd. The operators in \eq{LeffMD} are  also $P$ odd, and therefore overall $CP$ even. The operator in \eq{LeffED} is  $P$ even, and therefore overall $CP$ odd. The corresponding $CP$ properties  are hence as expected for being induced by a magnetic- and electric-dipole operator, respectively. 

The Lagrangians in \eq{LeffMD} and \eq{LeffED} are different from those (induced by anomalies)  studied in  Refs.~\cite{Keung:2008ve} or \cite{Dror:2017ehi}, from which  the  Landau-Yang theorem is bypassed by having massive final states.

\begin{figure}[t]
\begin{center}
\includegraphics[width=2.9in]{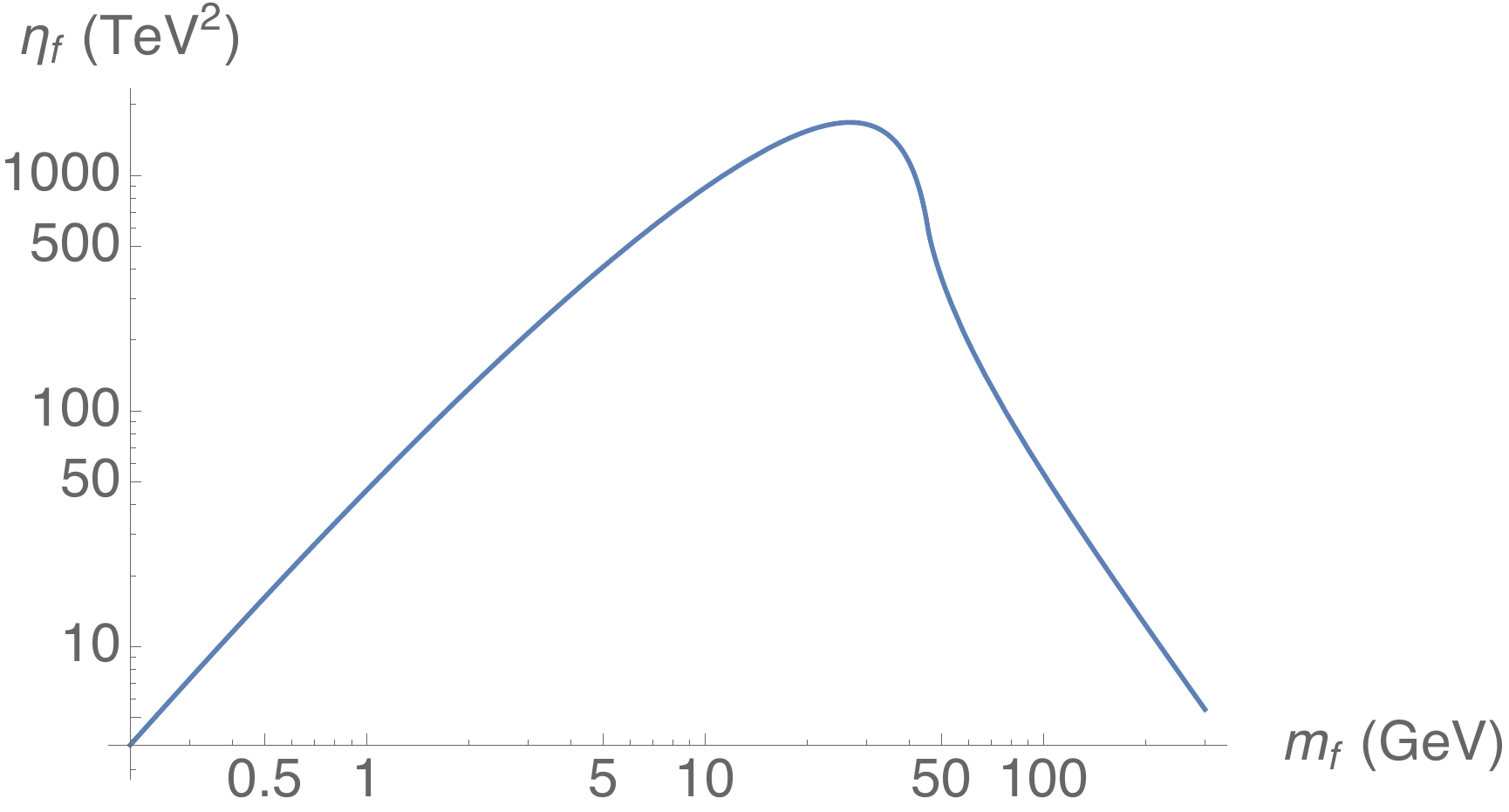}
\caption{\small Double logarithmic plot of the function \mbox{$\eta_f(m_f)=|\xi^f(m_f)|^2 (\Lambda/{\rm TeV})^2/r_f^2$} containing the loop fermion-mass  dependence characterizing  a single-fermion contribution to $\Gamma (Z\rightarrow \gamma \bar \gamma)$ in \eq{BRTOT}. The  nontrivial behavior around $m_f\sim 30$ GeV  explains the  flavor dependence of  $X_f$ in Table~\ref{BRs}.
\label{plotto} }
\end{center}
\end{figure}

\vskip0.3em 
\textit{Decay rate.}---The total $Z\to \gamma \bar \gamma$  decay width is obtained from \eq{sumM} and \eq{sumE}, and  is given by
\bea
\Gamma (Z\rightarrow \gamma \bar \gamma) &=&  \frac{ \alpha_{\scriptscriptstyle D} \alpha \, M_Z}
       {24\pi^3} \left(|C_{M}|^2 +  |C_{E}|^2\right) , \label{BRTOT}
\eea
where  $C_{M,E} = \sum_f d_{M,E}^f \, \xi^f (m_f)$,  and $\xi^f (m_f)$ is given in   \eq{CM}. In Fig.~\ref{plotto}, the function \mbox{$\eta_f(m_f)=|\xi^f(m_f)|^2 (\Lambda/{\rm TeV})^2/r_f^2$} [pinpointing the $m_f$ dependence of a single  $f$ SM-fermion  contribution to $\Gamma (Z\rightarrow \gamma \bar \gamma) \,$] is plotted versus the fermion mass.

The  branching ratios BR$_f=\Gamma_f/\Gamma_{Z}$, which by definition assume a single  
SM-fermion  contribution from flavor $f$ to \eq{BRTOT}, can be obtained from Table~\ref{BRs},
with BR$_f= \alpha_{\scriptscriptstyle D}  (|d_{M}^f|^2+|d_{E}^f|^2)/(\Lambda/{\rm TeV})^2 {X_f}$.
\begin{table}[t]
\small
\caption{BRs for the largest single-fermion $Z \gamma \bar \gamma$-loop contributions.   The BR corresponding to each SM fermion of flavor $f$ would be BR$_f= \alpha_{\scriptscriptstyle D}  (|d_{M}^f|^2+|d_{E}^f|^2)/(\Lambda/{\rm TeV})^2\; {X_f}$. The $CP$-conserving  and $CP$-odd parts are equal. 
 }
\begin{ruledtabular}
\begin{tabular}{|c|c|c|c|c|c|c|c|}
&$b$  &  $t$   &  $s$ &   $c$ & $\tau$ & $\mu$ & \cr
\hline
\rule{0.5em}{0pt} ${X_f}$  \rule{0.5em}{0pt}& 4.80 & 0.82  & 0.014 & 4.78 &  1.30 & 0.017 & $\!\!\!\!\times 10^{-9} $ \cr
\end{tabular}
\end{ruledtabular}   
\label{BRs}
\end{table}%
 For   lighter flavors ($u,d,e$)---whose coefficients  $d_M^f/\Lambda$ and   $d_E^f/\Lambda$ are strongly constrained by astrophysics~\cite{Hoffmann:1987et}---BR$_f$ turns out to be further suppressed by the loop $m_f$ dependence.
 
Assuming universal magnetic-  and electric-dipole moments, $d_M$ and  $d_E$, by summing up all fermion contributions (including  interferences)  for all heavier flavors in Table~\ref{BRs}, we obtain  
\bea
\!\!\!\!{\rm BR}(Z\rightarrow \gamma \bar \gamma)\!\!\!&\simeq& \!\!\!\! \frac{2.52  \;\alpha_{\scriptscriptstyle D }}{(\Lambda/{\rm TeV})^2}\; (|d_M|^2+|d_E|^2) \!\times \!10^{-8} \,. \label{BR}
\eea
The  resulting ${\rm BR}(Z\rightarrow \gamma \bar \gamma)$  is quite suppressed, resulting from an effective two-loop computation, also featuring a few partial cancellations.  
Depending on the values assumed for the dipole moments in \eq{ds}, the dark-sector energy scale $\Lambda$, and coupling $\alpha_{\scriptscriptstyle D}$, this channel could  be observable at colliders collecting  large samples of $Z$ bosons.

If we assume that the dipole momenta in \eq{ds} are produced at the limit for perturbative  interactions in the dark sector, where $g^f_L g^{f*}_R\simeq (4 \pi )^2$---we can take $d_E \simeq d_M \simeq 1/2$.  In this case, 
 assuming $\alpha_D$ of order 0.1 and $\Lambda$  around 1 TeV, one would get  
 ${\rm BR}(Z\rightarrow \gamma \bar \gamma)\simeq 10^{-9}$ from \eq{BR}.
 Smaller (and perhaps more realistic) couplings, like $d_E \simeq d_M \simeq 0.1$, 
 would anyhow give ${\rm BR}(Z\rightarrow \gamma \bar \gamma)\simeq  4 \times 10^{-11}$,
 for the same $\alpha_D$  and $\Lambda$.

Our prediction  stems from a weakly coupled UV complete model.
One could also envisage different frameworks with effective couplings generated, 
for  instance, by a nonperturbative dynamics, pushing the effecting $\Lambda$ scale to lower values, and correspondingly enhancing ${\rm BR}(Z\rightarrow \gamma \bar \gamma)$, and 
possibly saturating the present LEP limit of BR $< 10^{-6}$ on $Z$ monophoton decays~\cite{Acciarri:1997im}.

Note that while  the processes involving the dark-photon coupling 
to the Higgs in Ref.~\cite{Biswas:2016jsh} are mainly ruled by the $\alpha_D$ magnitude, 
the present $Z$-decay rate, similarly to the FCNC decays in Ref.~\cite{Gabrielli:2016cut}, is governed also by the $\Lambda$ scale parameter. 
The strongest bounds on the model mass spectrum,  and therefore on $\Lambda$,
come from messenger-pair production at the LHC. These 
(taking into account the possibility of different signatures) are similar to the current squarks and sleptons   bounds~\cite{Patrignani:2016xqp}.

\vskip0.3em 
\textit{Experimental perspectives.}---Large samples of $Z$ bosons produced in high-energy collisions will be needed to test the prediction in \eq{BR}. Present and future hadron colliders will collect a 
very large number of $Z$'s. On the other hand, separating the background for a final state involving  moderate missing
momentum  (like the typical final state in $Z\rightarrow \gamma \bar \gamma)$ would be in general very challenging in proton collisions. One could try to focus on boosted $Z$ systems,
which anyhow would deplete the statistics.
The cross section   for  $Z$-boson production at the LHC is  about $59 $ nb  (at 13 TeV)~\cite{Aad:2016naf}, which,  with an integrated luminosity of  300 fb$^{-1}$, would provide about $2 \times 10^{10}$ $Z$ bosons.  
Ten times more $Z$'s will be produced with the integrated luminosity of  3 ab$^{-1}$ expected at the HL-LHC. 
Assuming the dipole momenta in \eq{ds}  at the limit for perturbative  interactions in the dark sector, for $\alpha_D\simeq$  0.1 and $\Lambda\simeq$   1 TeV, corresponding to ${\rm BR}(Z\rightarrow \gamma \bar \gamma)\simeq 10^{-9}$,
one could have about 200 $Z\rightarrow \gamma \bar \gamma$ events at the
HL-LHC. A more favorable sample (with a less favorable environment) is expected 
at a 100 TeV collider---where the production cross section for the $Z$ boson is about  0.4 $\mu$b~\cite{Arkani-Hamed:2015vfh}, with the number of produced  $Z$'s around $10^{13}$ for a luminosity of 30 ab$^{-1}$, and $10^4$ $Z\rightarrow \gamma \bar \gamma$ decays.

The study of the final-state angular distribution for polarized $Z$ bosons  would allow us to distinguish the dark photon from the axion case~\cite{Dror:2017ehi}, the two having otherwise similar signatures.

The best opportunity to study the $Z\rightarrow \gamma \bar \gamma$ decay channel---and possibly discover a dark photon---would come from the clean environment of the Future Circular Collider (FCC-ee), with its projected production of $10^{13}$ $Z$ bosons~\cite{Gomez-Ceballos:2013zzn}.



\end{document}